\documentclass[a4paper]{jpconf}
\usepackage{graphicx}

\def\be {\begin{equation}}
\def\ee {\end{equation}}

\begin{document}
\title{Heavy quark dynamics in QCD matter}

\author{S. K. Das$^{1}$, F. Scardina $^{1,2}$, S. Plumari$^{1,2}$ and V. Greco$^{1,2}$}

\address{$^1$ Department of Physics and Astronomy, University of Catania, Via S. Sofia 64, I-
95125 Catania, Italy}

\address{$^2$ Laboratori Nazionali del Sud, INFN-LNS, Via S. Sofia 62, I-95123 Catania, Italy}

\ead{santosh@lns.infn.it}

\begin{abstract}
Simultaneous description of  heavy quark nuclear suppression factor $R_{AA}$ and the elliptic flow $v_2$ 
is a top  challenge for all the existing models.  We  highlight how  the temperature dependence of the energy 
loss/transport coefficients is responsible to address a large part of such a puzzle along with the the full 
solution of the Boltzmann collision integral for the momentum evolution of heavy quark. We consider four different 
models to evaluate the temperature dependence of drag coefficients of the heavy quark in the QGP. We have also 
highlighted the heavy quark dynamics in the presence of an external electromagnetic field which develops a sizable 
heavy quark directed flow, $v_1(y)$, can be measurable at LHC. 

\end{abstract}

\section{Introduction}
On going experimental efforts at Relativistic Heavy Ion Collider (RHIC) and Large Hadron Collider (LHC) energies is aimed at to create and characterized 
the properties of quark gluon plasma (QGP). The heavy mesons (mesons which contain one heavy quark, mainly c and b ) 
constitute a novel probe of the QGP properties, because they are produced in the early stage
of the heavy-ion collisions and they are the witness to the entire space-time evolution of the QGP. 

The two main experimental observables related to heavy quarks dynamics in QGP are: (i)  the nuclear suppression 
factor, $R_{AA}$~\cite{stare,phenixelat} which is the ratio of
the  $p_T$ spectra of heavy flavored mesons (D and B) produced 
in nucleus + nucleus collisions to those produced in proton + proton collisions scaled with the number of binary collision
and (ii) the elliptic flow, $v_2=\langle cos(2\phi_p)\rangle$~\cite{phenixelat,alicev2}, which is a measure of 
the anisotropy in the angular distribution of particle production.
Several theoretical efforts have been made to study the $R_{AA}$ and the $v_2$ measured in experiments 
within different models~\cite{rappv2,rappprl,Das,alberico,bass,Das:2015ana,gossiauxv2,gre,fs,Song:2015sfa,fs2,Cao:2016gvr, Younus:2013rja, Tripathy:2016hlg} (see also 
for heavy baryons~\cite{Das:2016llg}).
However all the approaches shown some difficulties to describe both the $R_{AA}$ and $v_2$ 
simultaneously.

\section{Results}
To address the  simultaneous description of $R_{AA}$ and $v_2$, we need to focus on the time evolution 
of $R_{AA}$ and $v_2$ to know how they develop during the expansion of the QGP. As shown in Ref.~\cite{Das:2015hla}, the 
$R_{AA}$ develops mostly at the very early stage of the QGP evolution  within 3-4 fm/c,
while the $v_2$ builds up later and consequently can be transferred to the heavy quarks also only at later stages when the 
fireball has reached temperatures close to the critical temperature. This indicates a more liquid-like behavior of 
the quark-gluon plasma and the T-dependence of the drag coefficient plays a significance role 
for a simultaneous description of $R_{AA}$ and $v_2$  as they are sensitive to the two different stages 
of the QGP evolution ($T_i$ and $T_c$). To investigate the influence of the temperature dependence of the drag (and diffusion) 
coefficients on heavy quark observables, four different models have been used to calculate the drag and diffusion coefficients.

Model-I (pQCD): In this case, the elastic collisions of heavy quarks with the bulk consist of light quarks, anti-quarks and gluons have been considered within
the framework of pQCD having temperature dependence of the coupling~\cite{zantow}:
\be
g^{-2}(T)=2\beta_0 ln\left(\frac{2\pi T}{a\,T_c ×}\right)+\frac{\beta_1}{\beta_0×}ln\left[ln\left(\frac{2\pi T}{a\,T_c ×}\right)\right]
\ee
where $\beta_0=(11-2N_f/3)/16\pi^2$, $\beta_1=(102-38N_f/3)/(16\pi^2)^2$ and $a=1.3$.
$N_f$ is the number of flavor and $T_C$ is the transition temperature. \\

Model-II (AdS/CFT): In this second case, we consider the drag force from the gauge/string duality 
i.e. AdS/CFT~\cite{Maldacena:1997re}, $\Gamma_{conf}= C \frac{T_{QCD}^2}{M_{HQ}}$, where $C=2.1\pm0.5$ and 
the diffusions coefficient deduced from fluctuations-dissipation~\cite{ali}. \\
Model-III (QPM):
In this case, we employ a quasi particle model (QPM)~\cite{salvo,vc} with T-dependent quasi-particle 
masses, $m_q=1/3g^2T^2$, $m_g=3/4g^2T^2$, along with
a T-dependent background field known as bag constant, tuned to reproduce the thermodynamics of 
the lattice QCD. The fit lead to the coupling, $g^2(T)=\frac{48\pi^2}{[(11N_c-2N_f)ln[\lambda(\frac{T}{T_c}-\frac{T_s}{T_c})]^2}$,   
where $\lambda$=2.6 and $T/T_s$=0.57. \\
Model-IV ($\alpha_{QPM}(T),m_q=m_g=0$):  
In this fourth case, we consider a model where the light quarks and gluons are massless but the 
coupling is taken from the QPM model discussed above. This case is merely a way to obtain a drag increasing toward $T_c$.   \\

\begin{figure}[ht]
\begin{minipage}{12pc}
\includegraphics[width=12pc, clip=true]{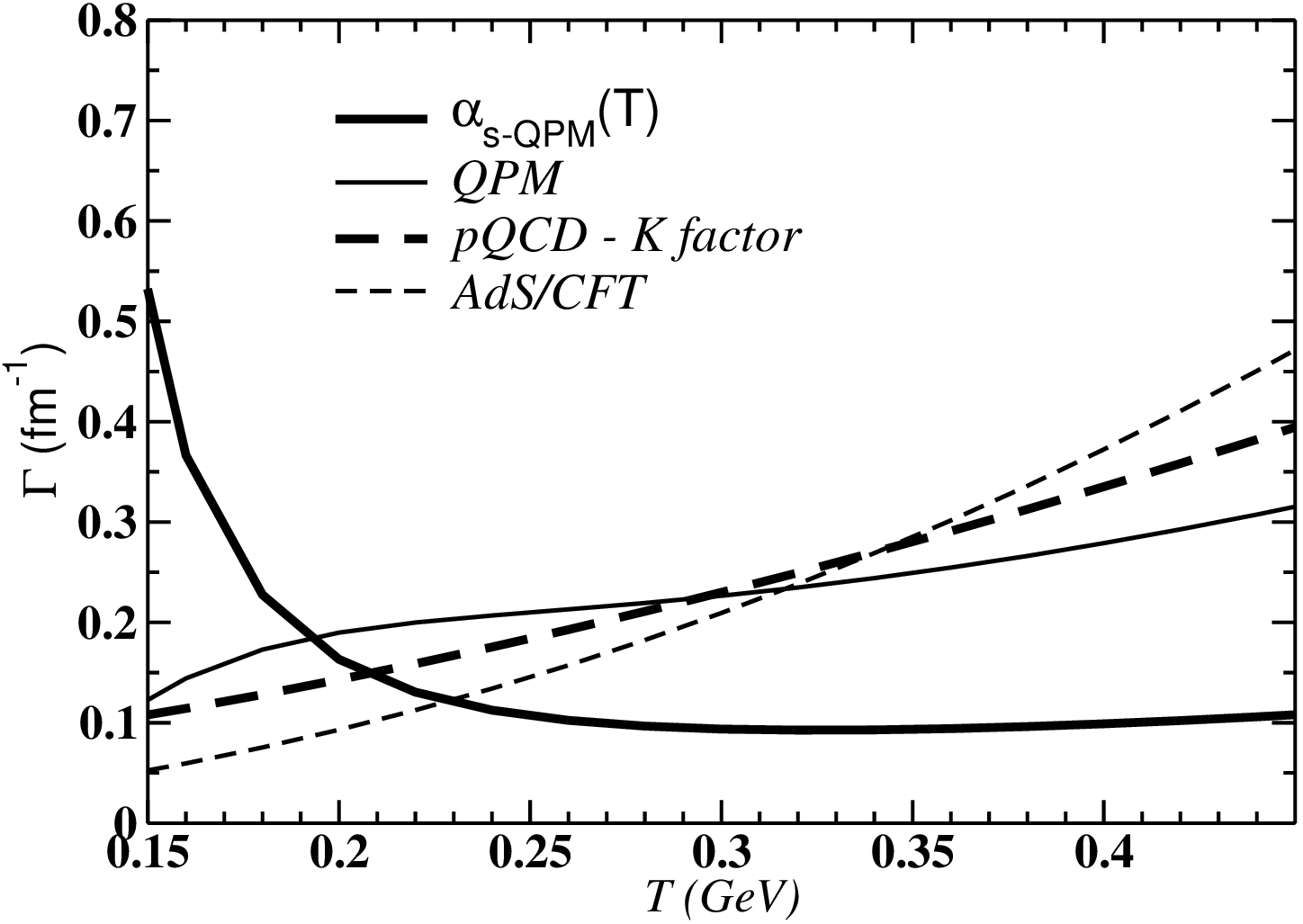}
\caption{Drag coefficients as a function of T at p=100 MeV.}
\label{drag}
\end{minipage}\hspace{1pc}%
\begin{minipage}{12pc}
\includegraphics[width=12pc, clip=true]{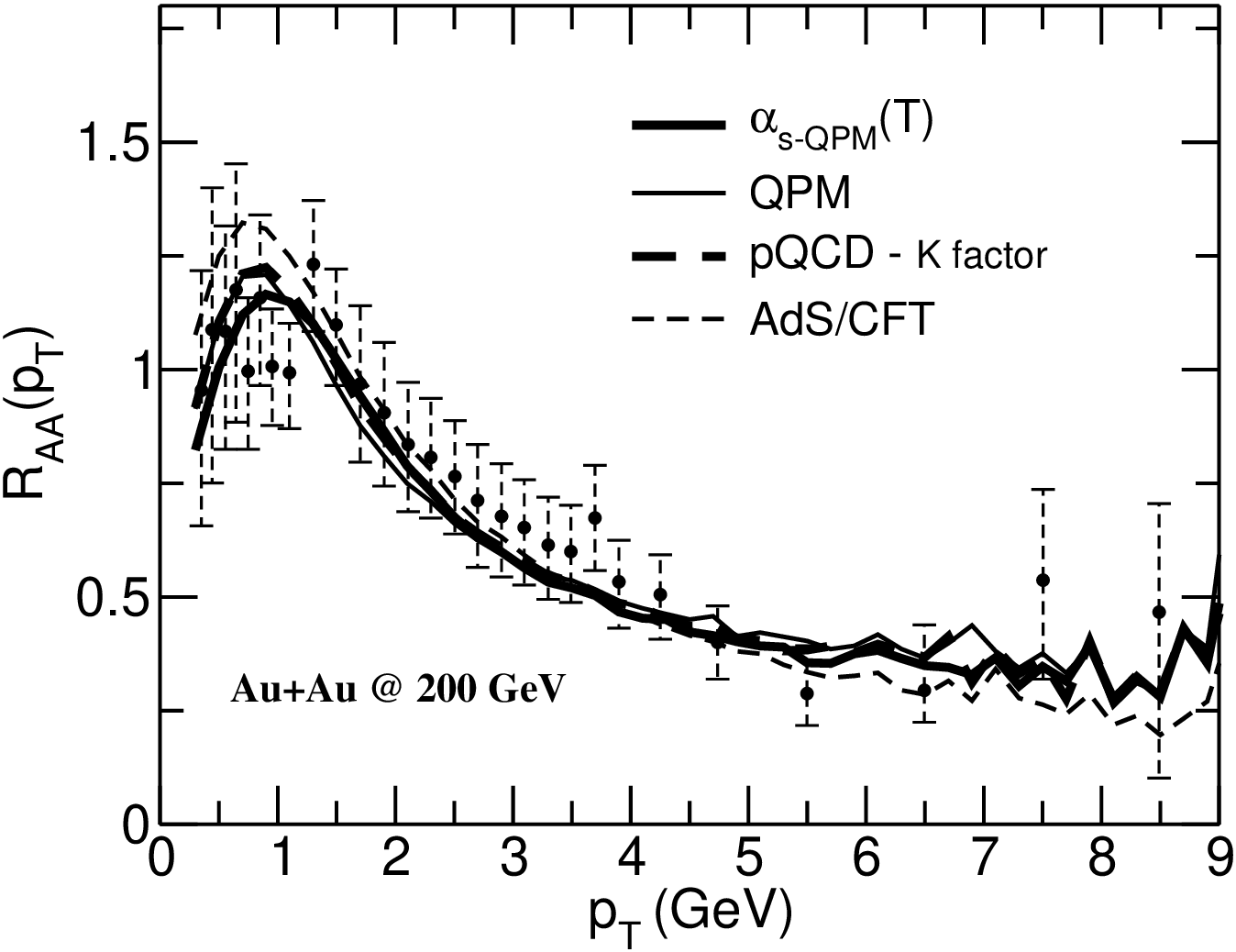}
\caption{ $R_{AA}$ as a function of $p_T$ for the minimum bias. }
\label{RAA_time1}
\end{minipage}\hspace{1pc}%
\begin{minipage}{12pc}
\includegraphics[width=12pc, clip=true]{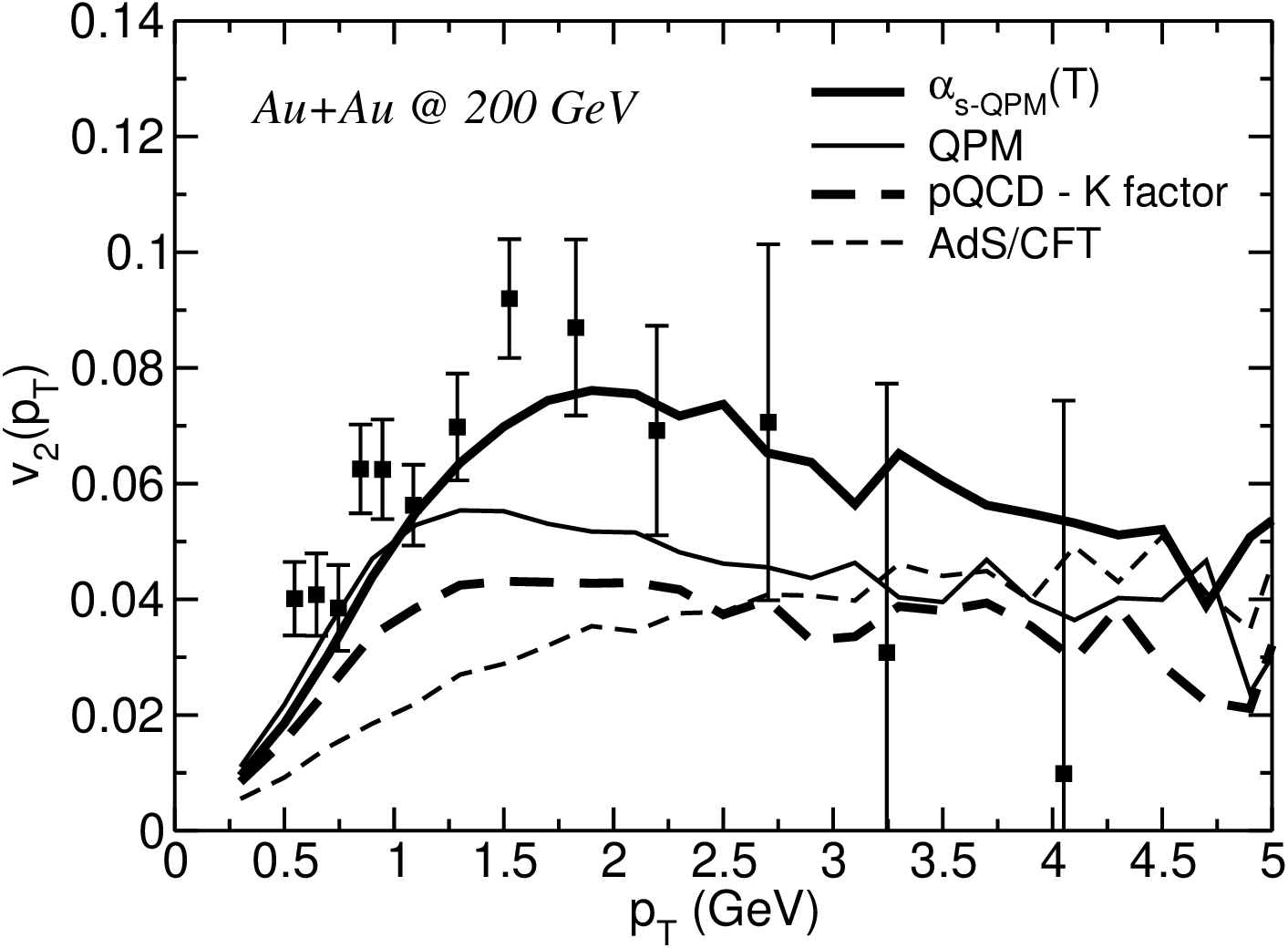}
\caption{ $v_2$ as a function of $p_T$ for the minimum bias. }
\label{v2_time1}
\end{minipage} 
\end{figure}

This fourth case has been considered to have  drag coefficient with a T dependence similar to
the T-matrix approach~\cite{rappprl,riek}. In Fig~\ref{drag}, we have shown 
the variation of the drag coefficients obtained within the four different model discussed above.
Our methodology is to reproduce the same $R_{AA}$ as of the experiments within all the four models, 
hence, these are the rescaled drag coefficients.

We have solved the Langevin dynamics to study the heavy quark momentum evolution in QGP 
starting from charm quark production in p-p collision~\cite{initial} as the initial charm quark distributions 
in the momentum space. To simulate the heavy quark dynamics in QGP, we need the bulk evolution.  
We are using the transport bulk which can reproduce the gross 
features of the bulk i.e the spectra and elliptic flow. For the $Au+Au$ collisions at the 
highest RHIC energy, we simulate the QGP evolution  
with initial temperature  in the center of the fireball is  $T_i=340$ MeV and the initial time  is $\tau_i=0.6$ fm/c. 
For the detail of the fireball evolution, we refer to our early works~\cite{Ruggieri:2013bda,Scardina:2012hy}.

In Fig~\ref{RAA_time1}, we have shown the variation of $R_{AA}$ 
as a function of $p_T$ for the four different models at RHIC energy. The $v_2$ 
for the same $R_{AA}$ has been plotted in Fig~\ref{v2_time1} for all the four models. 
It has been observed that for the same $R_{AA}$, the $v_2$ build-up can be quite different 
depending on the T-dependence of the drag coefficients. Larger the drag coefficient at $T_c$ larger 
is the $v_2$~\cite{Das:2015ana}. Similar effect has also been observed in the light quark sector~\cite{Liao:2008dk,Scardina:2010zz}. 
This implys, the correct temperature dependence of the drag coefficient has a vital role 
for a simultaneous description of heavy quark $R_{AA}$ and $v_2$.
The heavy quark observables are also sensitive to the hadronic rescattering~\cite{Das:2015hla,hp}, pre-equilibrium phase~\cite{Das:2015aga} 
as well as to the time evolution equation i.e where within Langevin or Boltzmann equation~\cite{fs}.
In  particular,  the  solution  of  the  full  two-body collision integral shows that the anisotropic 
flows are larger respect to those predicted by a Langevin dynamics.
For details we refer to our earlier works~\cite{fs}.

\begin{figure}[ht]
\begin{minipage}{17pc}
\includegraphics[width=12pc,clip=true]{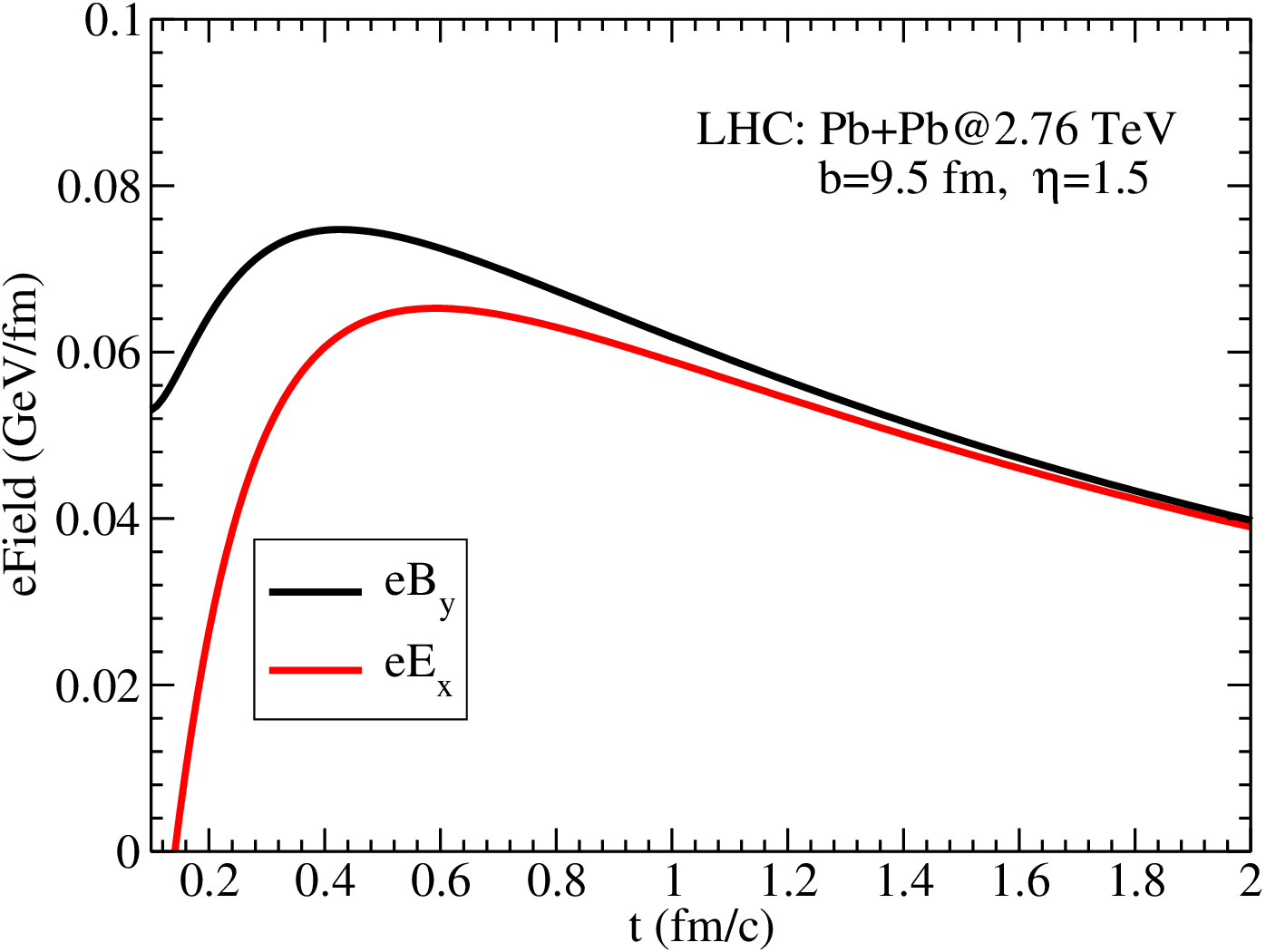}
\caption{Time evolution in 
the forward rapidity region of the magnetic field $eB_y$ and the electric field $eE_x$.}
\label{em}
\end{minipage}\hspace{2pc}%
\begin{minipage}{17pc}
\includegraphics[width=13pc, clip=true]{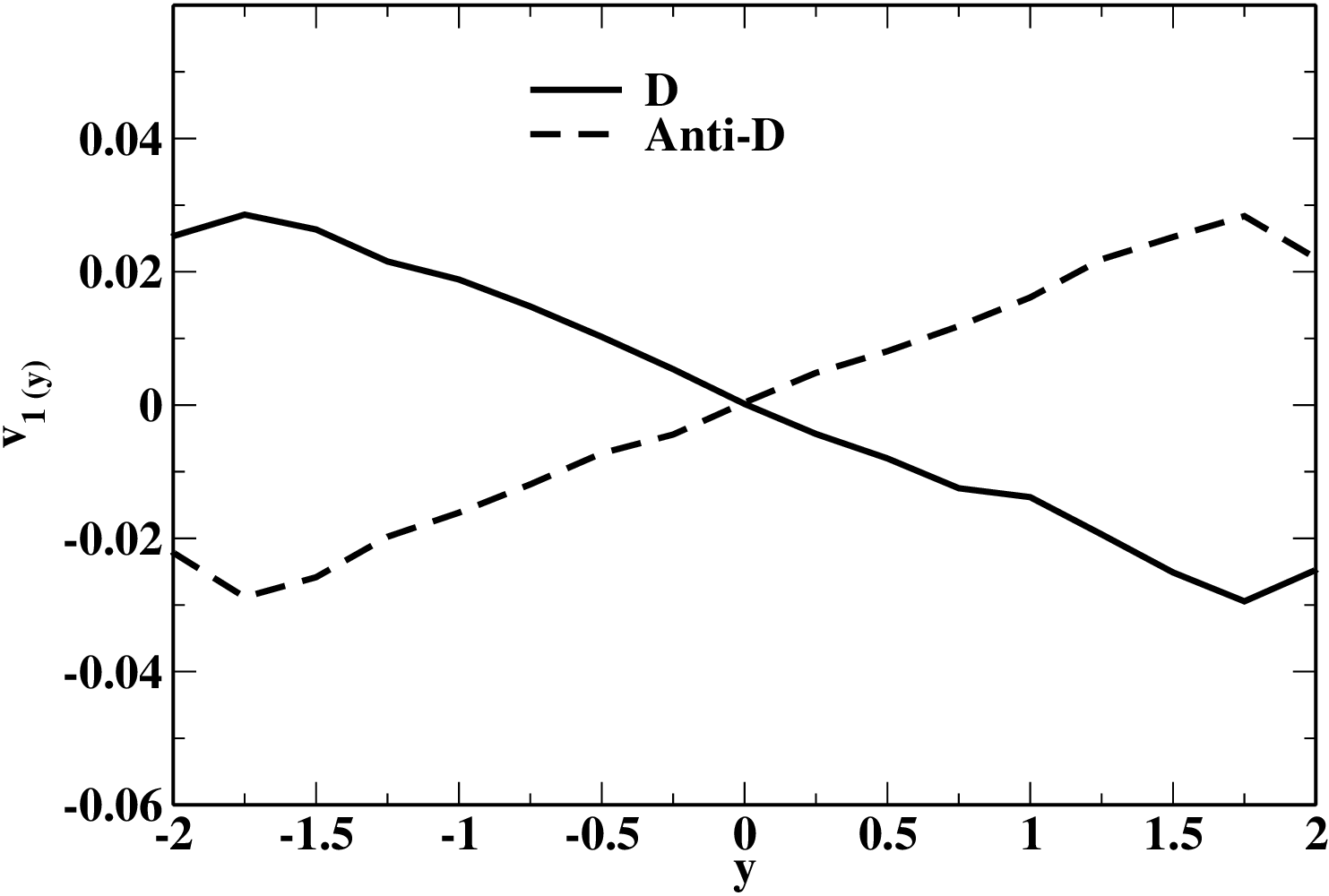}
\caption{Directed flow $v_1$ as a function of the rapidity in $Pb+Pb$ at $\sqrt{s}=2.76 \,\rm ATeV$ 
 for D meson [$c\overline q$] and anti-D meson [$\overline c q$].}
\label{v1}
\end{minipage}\hspace{2pc}%
\end{figure}

In the recent past it has been recognized that a very strong magnetic 
field~\cite{Kharzeev:2007jp,Fukushima:2008xe, Tuchin:2010vs} 
is created at early times in high energy heavy ion collisions.  
Since  heavy quarks are produced at the very early stage of evolution due to their large mass their 
dynamics can be affected by such a strong magnetic field~\cite{Fukushima:2015wck,Finazzo:2016mhm}.
The $\vec B$ field generated in non-central heavy ion collisions dominated by the component along the $\vec y$
axis, so its main effect is the induction of a current in the $xz$ plane.
On the other hand the time dependence of $\vec B$ generates  a electric field by Faraday's law, 
$\vec{\nabla} \cdot \vec E= \partial \vec B/\partial t $ which induces a current. For the calculation of
the electromagnetic field generated in ultra-relativistic heavy-ion collision, 
we refer  to the space-time solution developed in Ref.~\cite{Gursoy:2014aka}.  
In Fig.\ref{em} we show the time  dependence of the $B_y$ 
and electric field $E_x$  at finite space rapidity $\eta=1.5$ 
in a $Pb+Pb$ at $\sqrt{s}=2.76\,\rm ATeV$ at $b=9.5 \, \rm fm $ for a medium of $\sigma_{el}=0.023 \, \rm fm^{-1}$.
The impact of the electromagnetic field has been taken into account through the Lorentz force
as the external force in the Langevin equation. For the detail we refer to Ref.~\cite{Das:2016cwd}.

Magnetic field introduce a anisotropy which eventually leads to a substantial directed 
flow, $v_1=\langle p_x/p_T\rangle$, can be measurable at the experiments.
In Fig. \ref{v1} we have presented the resulting directed flow $v_1$ as a function of rapidity for charm
(black solid line) and anti-charm quarks (dashed line line). We find  a substantial $v_1$
at finite rapidity with a peak at $y\simeq \, 1.75$. The directed flow is negative for negative charged
particle (charm) at forward rapidity which means that the the magnetic field
dominates over the displacement induced by the Faraday current (generated due to the time dependence of
the magnetic field). This is a non trivial result and depends not only on the absolute magnitude of the magnetic field
but also on the strength  of heavy-light quark interaction. If we artificially increase the drag
coefficient by a factor 5  to achieve  a thermalization of the charm quarks similar to the light quark one,   
this will lead to a quite smaller $v1 (y)$.

\section{Summary and outlook}
In summary, we have presented how the $v_2$ build-up for the same $R_{AA}$ depends on the 
T-dependence of the interaction (drag and diffusion coefficients)  which is the key for a simultaneous description of heavy quark
$R_{AA}$ and $v_2$. We have also mentioned how the $v_2$ gets a boost from the 
Boltzmann dynamics by studying the heave quark momentum evolution and from the hadronic rescattering. 
We have also studied the heavy quark dynamics 
in the presence of electromagnetic field. Heavy quark develops a sizable directed flow $(v_1)$ 
which could be measurable at experiments. The present study suggests the heavy quark $v_1$ can be 
considered as a significant probe to characterize the magnetic field.

\section*{Acknowledgements} 
We acknowledge the support by the ERC StG under the QGPDyn
Grant n. 259684.

\end{document}